\documentclass[aps,prl,twocolumn,groupedaddress]{revtex4}
\usepackage{graphicx}
\usepackage{amsmath,amsfonts,amssymb}
\usepackage{color}
\usepackage{float}
\begin{document}

\title{ Local Scale Invariant Kaluza-Klein Reduction}
\author{Tomer Ygael}
\email{tomeryg@post.bgu.ac.il}
\author{Aharon Davidson}
\email{davidson@bgu.ac.il}
\homepage{http://www.bgu.ac.il/~davidson}
\affiliation{Physics Department, Ben-Gurion University
of the Negev, Beer-Sheva 84105, Israel}

\date{\today}

\begin{abstract}
 We perform the 4-dimensional Kaluza-Klein (KK)
reduction of the 5-dimensional locally scale invariant
Weyl-Dirac gravity.
While compactification unavoidably introduces an
explicit length scale into the theory, it does it in such
a way that the  KK radius can be integrated out from
the low energy regime, leaving the KK vacuum to still
enjoy local scale invariance at the classical level.
Imitating a $U(1)\times\tilde{U}(1)$ gauge theory, the
emerging 4D theory is characterized by a kinetic
Maxwell-Weyl mixing whose diagonalization procedure
is carried out in detail.
In particular, we identify the unique linear combination
which defines the 4D Weyl vector, and fully classify the
4D scalar sector.
The later consists of (using Weyl language) a co-scalar
and two in-scalars.
The analysis is performed for a general KK $m$-ansatz,
parametrized by the power $m$ of the scalar field which
factorizes the 4D metric. 
The no-ghost requirement, for example, is met provided
$-\frac{1}{2}\leq m \leq 0$.
An $m$-dependent dictionary is then established between
the original 5D Brans-Dicke parameter $\omega_5$ and
the resulting 4D $\omega_4$.
The critical $\omega_5=-\frac{4}{3}$ is consistently mapped
into critical $\omega_4 = -\frac{3}{2}$.
The KK reduced Maxwell-Weyl kinetic mixing cannot be scaled
away as it is mediated by a 4D in-scalar (residing within the
5D Weyl vector).
The mixing is explicitly demonstrated within the Einstein frame
for the special physically motivated choice of $m=-\frac{1}{3}$.
For instance, a super critical Brans-Dicke parameter induces
a tiny positive contribution to the original (if introduced via the
5-dimensional scalar potential) cosmological constant.
Finally, some no-scale quantum cosmological aspects are studied
at the universal mini-superspace level.
\end{abstract}


\maketitle
\section{Introduction}
The idea of local scale invariance theories, which is over 100 years old, has been studied in many physical contexts \cite{CT1,CT2,CT3,CT4}. Lately it has, once again, began to be the subject of intensive debate \cite{DB1,DB3,DB6,DB7,DB8,DB9}.
Only a few years after the introduction of general relativity, Weyl attempted the unification of electromagnetism and
gravity by the introduction of a \textbf{local} "gauge" transformation ($\Omega$ is coordinate dependent) under which the metric 
and the electromagnetic field would be jointly transformed,
\begin{equation}
\label{lst}
g_{\mu \nu} \rightarrow e^{2\Omega(x)} g_{\mu \nu} \quad ; \quad 
K_{\mu}(x) \rightarrow K_{\mu}(x) - \partial_{\mu}\Omega(x),
\end{equation}
with gravitation and electromagnetism thus unified by sharing a common $\Omega(x)$. In the course of developing his
theory Weyl also discovered the so-called Weyl tensor which, under the transformation (\ref{lst}), has the transformation rule
\begin{equation}
C^{\lambda}_{\mu \nu \sigma} \rightarrow C^{\lambda}_{\mu \nu \sigma}
\end{equation}
where all derivatives of $\Omega(x)$ drop out identically. The great appeal of this conformal symmetry is that its imposition
actually leads to a unique choice of gravitational action, namely the Weyl action 
\begin{equation}
{\cal S}_W = - \zeta_g \int C^{\lambda \mu \nu \sigma}
C_{\lambda \mu \nu \sigma} \sqrt{-g}d^4x
\end{equation}
where $\zeta_g$ has to be dimensionless. Thus conformal gravity possess a dimensionless coupling constant and forbids
a presence of any fundamental cosmological term, which will arise from the symmetry breaking of the theory. The field equations
of the theory yield a Schwarzschild-like metric with the addition of a linear term, which is suggested to be a solution to the
galactic rotation curve problem \cite{C21,C22,C23,C24}. Moreover it was suggested that by adding this term to a 'general' Standard 
Model the theory becomes "renormalizable" \cite{Hooft,renorm}.\\
Once scalar-tensor theories enter the game ${\cal S}_W$ is no longer unique. The simplest type of scalar-tensor theory is the Brans-Dicke theory \cite{BD}
described by the action
\begin{equation}
{\cal S}_{BD} = \int \left(\phi^2 R - 4 \omega g^{\mu \nu} \phi_{;\mu}\phi_{;\nu}\right)\sqrt{-g}d^4x ~.
\end{equation}
In addition to fully enjoying a global scale invariance ($\Omega$ in this case is constant) i.e.
\begin{equation}
g_{\mu \nu} \rightarrow e^{2\Omega} g_{\mu \nu} \quad ; \quad \phi
 \rightarrow e^{-\Omega} \phi
\end{equation}
this action may also enjoy a local scale symmetry under the transformation (\ref{lst}). This is only possible for the critical $\omega_c=-\frac{3}{2}$ case, where all $\Omega(x)$
derivatives drop out identically. Unfortunately this poses an issue as this critical case yields a kinetic "ghost" term which is nonphysical. Many of the recent local scale invariant 
theories involve the combining of a Brans-Dicke type Lagrangian with some standard model Lagrangian \cite{DB2,DB4,DB5,DB10,3}. Scale-invariant theories are attracting increased interest
due the strong physical motivation \cite{2,41,42} found in low energy particle physics. One example being the classical action of the standard model if the Higgs mass term 
is dropped. This invited the idea that the mass term may emerge from the vacuum expectation value of an additional scalar field \cite{BST}.\\
All the above make it apparent that local scale symmetry should be a fundamental symmetry in nature. Therefore we will show how it is possible to build a local scale symmetric theory with an arbitrary Brans-Dicke $\omega$
by using the Weyl-Dirac action for an arbitrary dimension. Starting with a 5-dimensional Weyl-Dirac action we make a Kaluza-Klein type reduction \cite{conKK} to 4-dimensions, and the resulting action is 
a local scale invariant theory which includes two scale-less scalar fields and one field with the proper length units. At the mini-superspace level the no-scale $C^2$ conformal cosmology is empty,
this encourages the use of the Weyl-Dirac cosmology and two scalar gravity-anti-gravity cosmology \cite{GaG1,GaG2,GaG3,QC}.\\
The latter local symmetry is translated into an additional constraint (on top of the Hamiltonian constraint). This allows the associated no-scale wave function of the Universe to solely depend on the scales-less fields. Near the Big Bang the wave function behavior is then governed by a scale-less scalar field \cite{conBB} which is a remanent of the 5D local scale symmetry.

\section{N-dim local scale symmetric gravity}
Our starting point is an n-dimensions Brans-Dicke-like action,
\begin{equation}
\label{action}
{\cal S} = \int \left(\phi^2 R - 4 \omega g^{\mu \nu} \phi_{;\mu}\phi_{;\nu}\right)\sqrt{-g}d^nx ~.
\end{equation}
Where $R$ is the N-dim Ricci scalar. As is the case in 4D, this n-dimensional action fully enjoys a global scale invariance. Additionally
for the critical case, with the problematic "ghost" term, this action enjoys a local scale invariance.
However it is possible to overcome the issue of the "ghost" term. By utilizing Weyl's geometry, it is
possible to achieve local scale invariance even for an arbitrary $\omega$.
The basic idea is to convert all tensors into co-tensors (denoted by the "star"
notation) by supplementing them with the Weyl vector field $K_{\mu}$, with the
transformation law
\begin{equation}
K_{\mu}(x) \rightarrow K_{\mu}(x) - \partial_{\mu}\Omega(x),
\end{equation}
and its divergence. A co-tensor has the transformation law
\begin{equation}
Y^*\rightarrow e^{p\Omega}Y^*,
\end{equation}
where $p$ is called the power of the co-tensor. We will denote square 
brackets as the power of the co-tensors, i.e., $\left[Y^*\right]=p$. In the case of $\left[Y^*\right]=0$ we say
that the co-tensor is an in-tensor. In n-dimensions the powers of the metric $g_{\mu \nu}$ and its inverse $g^{\mu \nu}$ are
\begin{equation}
\left[g_{\mu \nu}\right] = 2 ~~\Longleftrightarrow~~ \left[g^{\mu \nu}\right] = -2 ~.
\end{equation}
Therefore the power of $\sqrt{-g}$ is
\begin{equation}
\left[\sqrt{-g}\right] = n ~.
\end{equation}
Turning to our action (\ref{action}), both curvature and the covariant
scalar field derivative must be upgraded to their star counterparts, making the action
scale invariant.
Following Dirac and Weyl we write,
\begin{equation}
\begin{array}{c}
\displaystyle{R^*_{\mu \nu}= R_{\mu \nu} + \frac{n-2}{2}\left( K_{\mu;\nu} K_{\nu;\mu} + 2K_{\mu}K_
{\nu}\right) \hspace{20pt}} \\[3pt]
\displaystyle{+ g_{\mu \nu} \left( K^{\mu}_{;\mu} - (n-2)K^{\mu} K_{\mu}\right) \hspace{40pt}}
\end{array}
\end{equation}
\begin{equation}
\begin{array}{c}
\displaystyle{R^*=g^{\mu \nu}R^*_{\mu \nu} \hspace{160pt}} \\[3pt]
\displaystyle{\hspace{4pt}= R + 2(n-1)K^{\mu}_{;\mu} - (n-1)(n-2)K^{\mu}K_{\mu} ~.}
\end{array}
\end{equation}
$R^*_{\mu \nu}$ is the Ricci co-tensor which is in an in-tensor. $R^*$ is the
Ricci co-scalar. This means that for our action to be an in-scalar there is a need for
a scalar field to be coupled to the starred curvature. The powers of the Ricci co-tensor
and the scalar field are
\begin{equation}
\left[R^*\right] = -2 ~~ \left[\phi\right] = (2-n)/2 ~.
\end{equation}
We summarize the transformation laws for all the above,
\begin{eqnarray}
&R^*\rightarrow e^{-2\Omega}R^*,~ \sqrt{-g}\rightarrow
e^{n\Omega}\sqrt{-g},~ \phi \rightarrow e^{\frac{2-n}{2}\Omega}\phi&\\
&\Rightarrow ~\phi^2 R^*\sqrt{-g} \rightarrow \phi^2 R^*\sqrt{-g}& ~.
\end{eqnarray}
We now turn our attention to the scalar field kinetic term. The co-covariant 
derivative is defined as
\begin{equation}
\phi_{*\mu} = \phi_{;\mu} - \frac{2-n}{2}K_{\mu}\phi 
\end{equation}
with the transformation law
\begin{equation}
\phi_{* \mu} \rightarrow e^{\frac{2-n}{2}\Omega}\phi_{*\mu}
\end{equation}
similar to the scalar field. This means that the kinetic term is of the power, $\left[g^{\mu \nu}\phi_{*\mu}
\phi_{*\nu}\right]=-n$ (Remember: $\left[\sqrt{-g}\right]=n$). We can now write the local scale invariant n-dimensional
Brans-Dicke like action
\begin{equation}
\label{action*}
{\cal S} = \int \left(\phi^2 R^* - 4 \omega g^{\mu \nu} \phi_{*\mu}\phi_{*\nu}\right)\sqrt{-g}d^nx ~.
\end{equation}
A mandatory ingredient is a kinetic term for the Weyl vector field. Although, it is not directly required on plain
local scale symmetry grounds, in its absence $K_{\mu}$ will stay non-dynamical in nature. The transformation law (\ref{lst})
dictates the exact Maxwell structure, with the corresponding anti-symmetric differential 2-form given by
\begin{equation}
K_{\mu \nu} = K_{\mu;\nu} - K_{\nu; \mu} ~.
\end{equation}
It is a simple exercise to show that the Weyl vector field strength tensor power is $\left[K_{\mu \nu}\right] =0$. However the kinetic term's power is $\left[K^{\mu \nu}K_{\mu \nu}\right]=-4$ and as such must be coupled to the scalar field with the appropriate power of $p=(8-2n)/(2-n)$. Finally we can follow the Dirac prescription \cite{Dirac} and write the local scale invariant action in n dimensions
\begin{equation}
\begin{array}{c}
\displaystyle{	{\cal S} = \int \sqrt{-g}d^nx \Bigl(\phi^2 R^* - 4 \omega g^{\mu \nu} \phi_{*\mu}\phi_{*\nu}  }\\
\displaystyle{	  - \frac{1}{4}\phi^{\frac{8-2n}{2-n}}K^{\mu \nu}K_{\mu \nu}\Bigr)} ~.\hspace{30pt}
	\end{array}
\end{equation}
We explicitly write the starred part of the Lagrangian, 
\begin{equation}
\begin{array}{c}
\label{n-dim}
\phi^2 R^* - 4 \omega g^{\mu \nu} \phi_{*\mu}\phi_{*\nu} = \hspace{170pt}\\
\phi^2 R - 4 \omega g^{\mu \nu} \phi_{;\mu}\phi_{;\nu} + (2-n)\omega\left(K^{\mu}\phi^2\right)_{;\mu} \hspace{35pt} \\ [3pt]
 + ((n-1)+(n-2)\omega)\left(2K^{\mu}_{;\mu} - (n-2)K^{\mu}K_{\mu}\right)\phi^2   
\end{array}
\end{equation}
Notice the interesting choice of $\omega=\frac{1-n}{n-2}$. This yields, up to a total derivative, a local scale
invariant action without the need for Weyl's vector $K_{\mu}$, i.e.,
\begin{equation}
\phi^2 R^* - 4 \omega g^{\mu \nu} \phi_{*\mu}\phi_{*\nu} = \phi^2 R - 4 \omega g^{\mu \nu}
 \phi_{;\mu}\phi_{;\nu} ~.
\end{equation}
For $n=4$ we find that $\omega_4 = -\frac{3}{2}$. This is the known $\omega_ {BD}=-\frac{3}{2}$ which provides the local 
scale invariant Brans-Dicke action.

\section{5-dimensional Weyl-Dirac theory}
In five dimensions the Weyl-Dirac prescription reads,
\begin{equation}
\begin{array}{c}
	\displaystyle{{\cal S} = \int d^5x \sqrt{-G} \left(\phi^2 \hat{R}^* - 4 \omega_5 G^{MN} \phi_{*M}\phi_{*N} \right.} \\
	 \displaystyle{\left.- \frac{1}{4}\phi^
{\frac{2}{3}}
K^{MN}K_{MN} + V(\phi)\right)} ~. \hspace{30pt}
	\end{array}
\end{equation}
$G^{MN}$ is the 5D metric (upper case letters denoting 5D coordinates), and $\hat{R}^*$ is the, 5D, co-covariant Ricci scalar
\begin{equation}
\hat{R}^*= \hat{R} + 8 K^{M}_{;M} - 12 K^{M}K_{M} ~.
\end{equation}
$\hat{R}$ denotes the 5D Ricci scalar. Furthermore $\phi_{*M}$ is the the 5D co-covariant derivative defined by
\begin{equation}
\phi_{*M} = \phi_{;M} + \frac{3}{2}K_{M}\phi ~.
\end{equation}
We note that the scalar potential term in the action must be of a specific form, $~V(\phi)=-2\Lambda \phi^{10/3}$, to keep our action an in-action. However if one wishes to explicitly break the Weyl symmetry, changing the potential is the simplest way this can be done. On pedagogical grounds we will continue with a
general $V(\phi)$. Note that all covariant derivatives (i.e. $K^M_{;M}$) are the 5D variants and will be so for the rest of this section. Altogether, up to the total derivative $(2-n)\omega\left(K^{\mu}\phi^2\right)_{;\mu}$ and a full re-arrangement of the various terms floating around, the non-critical (arbitrary $\omega$)
local Weyl invariant theory can be described in a somewhat more familiar language
\begin{equation}
\begin{array}{c}
	\displaystyle{{\cal S} = \int d^5x  \sqrt{-G} \Bigl[ \phi^2 \hat{R} - 4 \omega_5 G^{MN} \phi_{;M}\phi_{;N} + V(\phi) \hspace{170pt}}  \\[3pt]
	\displaystyle{+ (4+3\omega_5)\left(2K^M_{;M} - 3K^MK_M\right)\phi^2  - \frac{1}{4} \phi^{\frac{2}{3}}K^{MN}K_{MN} \Bigr] ~.\hspace{130pt}}
	\end{array}
\label{5dun}
\end{equation}
Associated with this, non-critical, Lagrangian and corresponding to variations with respect to $G_{MN}, \phi$ and $K_{\mu}$, respectively, are the following field equations:
\begin{subequations}
\begin{widetext}
\begin{flalign}
\phi^2\left(R_{MN} - \frac{1}{2}G_{MN}\hat{R}\right) = & \nonumber\\
&- \phi^2_{;MN} + G_{MN} (\phi^{2})^{;M}_{~;M} - 4 
\omega_5 \phi_{;M} \phi_{;N} + 2 \omega_5G_{MN} \phi^{;M}\phi_{;M} \nonumber \\
&- \frac{1}{2}(4+3\omega_5)G_{MN}\left(2K^M_{;M} - 3 K^MK_M\right)\phi^2 - 3(4+3\omega_5)\phi^2
K_M K_N \nonumber \\
 &- (4+3\omega_5)\left(K_N\phi^2_{;M} + K_{M}\phi^2_{;N}\right) + (4+3\omega_5)G_{MN}\left(K^M
 \phi^2 \right)_{;M} \nonumber \\
 & - \frac{1}{2}\phi^{\frac{2}{3}}K^A_MK_{NA} + \frac{1}{8}\phi^{\frac{2}{3}}G_{MN}K^{MN}K_{MN}
  + \frac{1}{2}G_{MN}V(\phi) ~,
\end{flalign}
\end{widetext}
\begin{equation}
\begin{array}{c}
 \omega_5G^{MN}\left(\phi^2_{;MN}- 2\phi_{;M}\phi_{;N}\right) = \hspace{190pt}\\[3pt]
  \displaystyle{-\phi^2 \hat{R} + \frac{1}{2}V'(\phi) +\frac{1}{12}\phi^{\frac{2}{3}}K^{MN}K_{MN}}\\ [3pt]
 \displaystyle{- (4+3\omega_5) \left( 2 K^M_{;M} - 3 K^N K_N\right)\phi^2
  } ~,
\end{array}
\end{equation}
\begin{equation}
\left(\phi^{\frac{2}{3}}K^{MN}\right)_{;N} = 2 \left(4+\omega_5\right)G^{MN}\left(\phi^2_{;N} + 
3\phi^2K_{N}\right).\\
\end{equation}
\end{subequations}
As the underlying symmetry has not been broken, as long as $V(\phi)$ allows it, the equations 
of motion are also scale invariant. This allows them to be written in their "starred" form via
some straight forward calculations. The starred variants of the 5D equations 
of motion for $G_{MN}, \phi$ and $K_{\mu}$ ,respectively, are:
\begin{subequations}
\begin{equation}
\begin{array}{c}
	\phi^2 \left( R^*_{MN} - \frac{1}{2}G_{MN} \hat{R}^*\right) = \hspace{170pt}\\[7pt]
	\hspace{30pt}\displaystyle{\phi^2_{*M*N} - G_{MN}(\phi^2)^{*A}_{~*A} + 4 \omega_5\phi_{*M}\phi_{*N}} \\ [3pt]
	 \hspace{40pt} \displaystyle{- 2\omega_5G_{MN}G^{PQ}\phi_{*P}\phi_{*Q} + \frac{1}{2}\phi^{\frac{2}{3}}K^A_M K_{NA}} \\[5pt]
	\hspace{30pt} \displaystyle{-\frac{1}{8}\phi^{\frac{2}{3}}G_{MN}K^{PQ}K_{PQ} + \frac{1}{2}G_{MN}V(\phi)} ~,
	\end{array}
\end{equation}
\begin{equation}
\begin{array}{c}
	2\omega_5 G^{MP} \phi^2_{*M*P} = \hspace{150pt}\\
	\hspace{40pt}- \phi^2 \hat{R}^* + 4\omega_5G^{PQ}\phi_{*P}\phi_{*Q} \\ [4pt]
	\hspace{55pt} + \frac{1}{12}\phi^{\frac{2}{3}}K^{MN}K_{MN} + \frac{1}{2}V'(\phi) ~,
	\end{array}
\end{equation}
\begin{equation}
\left( \phi^{\frac{2}{3}}K^{MP}\right)_{*P} = 2 (4 + 3 \omega_5)G^{MP}\phi^2_{*P} ~.
\end{equation}
\end{subequations}
It is note worthy to mention that although the "starred" equations are simpler in form than their
"unstarred" counter-parts, there is a price to pay. The staring of the equation removes our physical intuition of the problem. As such we must always perform our analysis in the more complicated "un-starred" version of the equations. In this paper we will not be analyzing the 5D equations of motion. Instead, we will make a Kluza-Klein type reduction of the 5D action, vary it with respect to the relevant fields, and obtain the equations of motion in 4D. It is possible to perform the Kaluza-Klien reduction on the 5D equations of
motion. However this will make the understanding of the embedded 4D theory more complicated.\\
\section{Transformations and the K-K ansatz}
Working in higher dimensions, although very interesting, does not necessarily give us any intuition 
about our 4D world. In order to find the embedded 4D theory we must reduce the action from
the higher dimension to 4D. This is done by decomposing the 5D elements into their 4D counter-parts. The decomposition is actually done in accordance to the coordinate transformation laws of the 5D
elements
\begin{subequations}
\begin{flalign}
&G'_{MN} = \frac{\partial x_A}{\partial x'_M} \frac{\partial x_B}{\partial x'_N}G_{AB} ~,\\
&K'_M = \frac{\partial x_A}{\partial x'_M}K_A ~,\\
&\phi' = \phi ~.
\end{flalign}
\end{subequations}
We denote the 5D elements with upper-case letter and the 4D elements with lower-case
letters. Next we single out the 5th dimension and re-write $G_{MN}$ coordinate transformation laws
\begin{widetext}
\begin{eqnarray}
&&G'_{\mu \nu} = \frac{\partial x_A}{\partial x'_{\mu}} \frac{\partial x_B}{\partial x'_{\nu}}G_{AB} = 
\frac{\partial x_a}{\partial x'_{\mu}} \frac{\partial x_b}{\partial x'_{\nu}}G_{ab} + \frac{\partial x_a}
{\partial x'_{\mu}} \frac{\partial x_5}{\partial x'_{\nu}}G_{a5}+ \frac{\partial x_5}{\partial x'_{\mu}} 
\frac{\partial x_b}{\partial x'_{\nu}}G_{5b} + \frac{\partial x_5}{\partial x'_{\mu}} \frac{\partial x_5}
{\partial x'_{\nu}}G_{55} ~,\\
&&G'_{\mu 5} = \frac{\partial x_A}{\partial x'_{\mu}} \frac{\partial x_B}{\partial x'_5}G_{AB} = 
\frac{\partial x_a}{\partial x'_{\mu}} \frac{\partial x_b}{\partial x'_5}G_{ab} + \frac{\partial x_a}
{\partial x'_{\mu}} \frac{\partial x_5}{\partial x'_5}G_{a5}+ \frac{\partial x_5}{\partial x'_{\mu}} 
\frac{\partial x_b}{\partial x'_5}G_{5b} + \frac{\partial x_5}{\partial x'_{\mu}} \frac{\partial x_5}
{\partial x'_5}G_{55} ~,\\
&&G'_{5 \nu} = \frac{\partial x_A}{\partial x'_5} \frac{\partial x_B}{\partial x'_{\nu}}G_{AB} = 
\frac{\partial x_a}{\partial x'_5} \frac{\partial x_b}{\partial x'_{\nu}}G_{ab} + \frac{\partial x_a}
{\partial x'_5} \frac{\partial x_5}{\partial x'_{\nu}}G_{a5}+ \frac{\partial x_5}{\partial x'_5} 
\frac{\partial x_b}{\partial x'_{\nu}}G_{5b} + \frac{\partial x_5}{\partial x'_5} \frac{\partial x_5}
{\partial x'_{\nu}}G_{55} ~,\\
&&G'_{5 5} = \frac{\partial x_A}{\partial x'_5} \frac{\partial x_B}{\partial x'_5}G_{AB} = 
\frac{\partial x_a}{\partial x'_5} \frac{\partial x_b}{\partial x'_5}G_{ab} + \frac{\partial x_a}
{\partial x'_5} \frac{\partial x_5}{\partial x'_5}G_{a5}+ \frac{\partial x_5}{\partial x'_5} 
\frac{\partial x_b}{\partial x'_5}G_{5b} + \frac{\partial x_5}{\partial x'_5} \frac{\partial x_5}{\partial x'_5} G_{55}~.
\end{eqnarray}
\end{widetext}
Additionaly we re-write $K_A$ coordinate transformation laws
\begin{eqnarray}
&&K'_{\mu} = \frac{\partial x_A}{\partial x'_{\mu}}K_A = \frac{\partial x_a}{\partial x'_{\mu}}K_a + 
\frac{\partial x_5}{\partial x'_{\mu}}K_5 ~,\\
&&K'_{5} = \frac{\partial x_A}{\partial x'_{5}}K_A = \frac{\partial x_a}{\partial x'_{5}}K_a + 
\frac{\partial x_5}{\partial x'_{5}}K_5 ~.
\end{eqnarray}
Lastely, the scalar field does not go undergo a coordinate transformation, i.e., $\phi' = \phi$. Furthermore,
using the Kaluza - Klein ansatz, for which the vacuum solely allows for $x_a(x'_{\mu})$ and $x_5 = x'_5+\Lambda(x'_{\mu})$
we can write the $G_{MN}$ coordinate transformation as follows
\begin{align}
\begin{array}{c}
	 \displaystyle{G'_{\mu \nu} = \frac{\partial x_a}{\partial x'_{\mu}} \frac{\partial x_b}{\partial x'_{\nu}}G_{ab} + 
\frac{\partial x_a}{\partial x'_{\mu}} \frac{\partial \Lambda}{\partial x'_{\nu}}G_{a5}} \\
	\hspace{25pt} \displaystyle{+ \frac{\partial
 \Lambda}{\partial x'_{\mu}} \frac{\partial x_b}{\partial x'_{\nu}}G_{5b}+ \frac{\partial \Lambda}{\partial
  x'_{\mu}} \frac{\partial \Lambda}{\partial x'_{\nu}}G_{55}} ~, 
	\end{array}
\end{align}
\begin{align}
G'_{\mu 5} =  \frac{\partial x_a}{\partial x'_{\mu}}G_{a5} + \frac{\partial \Lambda}{\partial x'_{\mu}}G_{55} ~,
\end{align}
\begin{align}
G'_{5 \nu} = \frac{\partial x_b}{\partial x'_{\nu}}G_{5b} + \frac{\partial \Lambda}{\partial x'_{\nu}}G_{55} ~,
\end{align}
\begin{equation}
G'_{5 5} = G_{55} ~.
\end{equation}
These are accompanied by,
\begin{eqnarray}
&&K'_{\mu} = \frac{\partial x_a}{\partial x'_{\mu}}K_a + \frac{\partial \Lambda}{\partial x'_{\mu}}K_5 ~,\\
&&K'_{5} = K_5 ~.
\end{eqnarray}
Finally the 4D elements, embedded in the 5D elements, are recovered. We identify the coordinate transformation laws of:
\begin{itemize}
\item The 4D metric
\begin{equation}
g'_{\mu \nu} = \frac{\partial x_a}{\partial x'_{\mu}}\frac{\partial x_b}{\partial x'_{\nu}}g_{ab}
\end{equation}
\item Two 4D vectors
\begin{equation}
\frac{G_{\mu 5} }{G_{ 55}} \rightarrow \frac{\partial x_a}{\partial x'_{\mu}}\frac{G_{a 5} }{G_{ 55}} + 
\frac{\partial \Lambda}{\partial x'_{\mu}}, \qquad \frac{K_{\mu}}{K_5}\rightarrow \frac{\partial x_a}
{\partial x'_{\mu}}\frac{K_a}{K_5} + \frac{\partial \Lambda}{\partial x'_{\mu}}.
\end{equation}
\item A scalar field - $G'_{5 5} = G_{55}$.
\end{itemize}
It is quite remarkable that the special combination $\frac{G_{\mu 5} }{G_{ 55}} - \frac{K_{\mu}}{K_5}$ is
in-fact gauge invariant. Moreover this combination will later play a key part in our 4D theory.\\
We may more conveniently show off the 4D embedding in the 5D elements by writing
\begin{equation}
		G_{MN} = S^{m+1}\left(
	\begin{array}{c|c}
  S^{-1}g_{\mu \nu} + A_{\mu}A_{\nu} & A_{\mu} \\
  \hline
  A_{\nu} & 1
 \end{array}\right) ~,
\end{equation}
\begin{equation}
K_{M} = s \left(
	\begin{array}{c}
	 V_{\mu} \\ 
     \hline
        1
 \end{array}\right) ~.
	\end{equation}
With $m$ being an arbitrary power of the scalar field. This, however, will not affect the underlying physics of the problem. Additionally it is possible to find a simple dictionary between different choices of $m$.\\
We start by choosing $S=\tilde{S}^p$ and $g_{\mu \nu} = \tilde{S}^q \tilde{g}_{\mu \nu}$ such that
\begin{equation}
\begin{array}{c}
		G_{MN} = \tilde{S}^{p m}\left(
	\begin{array}{c|c}
  \tilde{S}^{q}\tilde{g}_{\mu \nu} + \tilde{S}^{p} A_{\mu}A_{\nu} & \tilde{S}^{p} A_{\mu} \\
  \hline
  \tilde{S}^{p} A_{\nu} & \tilde{S}^p
 \end{array}\right) \hspace{70pt} \\[15pt]
 = \tilde{S}^{p (m+1)-1}\left(
	\begin{array}{c|c}
  \tilde{S}^{q-p+1}\tilde{g}_{\mu \nu} + \tilde{S} A_{\mu}A_{\nu} & \tilde{S} A_{\mu} \\
  \hline
  \tilde{S} A_{\nu} & \tilde{S}
 \end{array}\right). 
\end{array}
	\end{equation}
Wanting to preserve the original form we choose $q=p-1$
\begin{equation}
\tilde{S}^{p (m+1)}\left(
	\begin{array}{c|c}
 \tilde{S}^{-1} \tilde{g}_{\mu \nu} +  A_{\mu}A_{\nu} &  A_{\mu} \\
  \hline
   A_{\nu} & 1
 \end{array}\right).
\end{equation}
Thus a simple dictionary is found between the different $m$,
\begin{equation}
\tilde{m} = p(m+1)-1.
\end{equation}
Note that the case of $m\neq -1$ is exceptional and cannot be transformed. In this case the action exclusively
enjoys an extra local scale symmetry. This yields a $K_{\mu}$ independent local scale invariant action i.e the Brans-
Dicke action with $\omega_{BD}=-3/2$. We will continue our analysis under the assumption that $m\neq-1$.\\
With the 4-vectors in mind we check how the original 5D local scale symmetry reflects on the 4D theory.
Recalling our action (\ref{action}) the 5D the transformation laws are,
\begin{equation}
G_{MN} \rightarrow e^{2\Omega}G_{MN}, \quad K_M\rightarrow K_M + \Omega_{;M}, \quad \phi\rightarrow e^{-\frac
{3\Omega}{2}}\phi.
\end{equation}
Writing the metric explicitly,
\begin{eqnarray}
		e^{2\Omega} G_{MN} =e^{2\Omega} S^{m+1}\left(
	\begin{array}{c|c}
  S^{-1}g_{\mu \nu} + A_{\mu}A_{\nu} & A_{\mu} \\
  \hline
  A_{\nu} & 1
 \end{array}\right),
\end{eqnarray}
we deduce the transformation laws of the 4D elements which constitute the 5D metric,
\begin{equation}
\begin{array}{c}
	S^{m+1}\rightarrow e^{2\Omega} S^{m+1} \Rightarrow S^m \rightarrow e^{\frac{2m\Omega}{m+1}}S^{m} ~,\\
	A_{\mu}\rightarrow A_{\mu}~, \quad g_{\mu \nu} \rightarrow e^{2 \tilde{\Omega}}g_{\mu \nu} ~,
\end{array}
\end{equation}
where we have denoted $\Omega$ as the 5D transformation parameter and $\tilde{\Omega}$ as the the 4D transformation parameter. Additionally we emphasis that only $\Omega(x_{\mu})$ is allowed in-order for the 4D theory to stay local scale invariant. This is owed to the fact that for the non-vacuum case all $x_5$ dependent fields need to be Fourier-expanded, including $\Omega$. It is a simple exercise to show that
once $\Omega$ is Fourier-expanded along side the rest of the fields the transformation laws are no longer defined by a single $\Omega$. In this case the action does not remain local scale invariant and must be modified accordingly if one wishes for it to remain local scale invariant.\\
Since we know how $S^m$ and  $g_{\mu \nu}$ transform both individually and as a product we can deduce that
\begin{equation}
S^m g_{\mu \nu} \rightarrow e^{\frac{2m\Omega}{m+1}+2\tilde{\Omega}}S^{m} g_{\mu \nu} \quad \Rightarrow \quad 
\Omega = \frac{m\Omega}{m+1}+\tilde{\Omega} ~.
\end{equation}
Hence the relation between the 4D and 5D transformation parameters is
\begin{equation}
\tilde{\Omega} = \frac{\Omega}{m+1} ~. 
\end{equation}
Next we write the 5D transformation of the Weyl gauge vector in detail,
\begin{equation}
K_{M} \rightarrow K_M + \Omega_{;M} =  s \left(
	\begin{array}{c}
	 V_{\mu} \\ 
     \hline
        1
 \end{array}\right) +  \left(
	\begin{array}{c}
	 \Omega_{;\mu} \\ 
     \hline
        0
 \end{array}\right) ~.
\end{equation}
Using the above relations we find the 4D transformation rules
\begin{equation}
\begin{array}{ccc}
	V_{\mu}&\rightarrow& V_{\mu} + \frac{(m+1)}{s}\tilde{\Omega}_{,\mu} \\[5pt]
	s&\rightarrow ~~s~. &
\end{array}
\end{equation}
Finally the scalar field 4D transformation is trivially given by $\phi \rightarrow e^{-\frac{3}{2}(m+1)\tilde{\Omega}}\phi$.\\
To summarize, after the reduction our 4D fields will have the following scale transformation laws
\begin{equation}
\begin{array}{c}
	g_{\mu \nu} \rightarrow e^{2 \tilde{\Omega}}g_{\mu \nu}~, ~ A_{\mu} \rightarrow A_{\mu}~, S \rightarrow e^{2\tilde{\Omega}}S ~,\\[5pt]
	V_{\mu} \rightarrow V_{\mu} + \frac{(m+1)}{s}\tilde{\Omega}_{;\mu}~, ~ s\rightarrow s~, \phi \rightarrow e^{\frac{3}{2}(m+1)\tilde{\Omega}}\phi ~.
\end{array}
\end{equation}
However, the theory does not yet have a 4D Weyl gauge vector. Although $V_{\mu}$ might appear as a good candidate it does not fill the requirements for the role. Unlike $V_{\mu}$ the Weyl vector is gauge invariant. Consequently we recall the special combination
\begin{equation}
W_{\mu} \equiv s\left(V_{\mu} - A_{\mu}\right),
\end{equation}
which is also gauge invariant. Additionally its scale transformation is 
\begin{equation}
W_{\mu} \equiv s\left(V_{\mu} - A_{\mu}\right) \rightarrow s\left(V_{\mu} - A_{\mu}\right) + (m+1) \tilde{\Omega}_{; \mu},
\end{equation}
which is very similar to the transformation law of the Weyl gauge vector. In a way this may be called the "un-normalized" 4D Weyl gauge vector. In our given case of $m\neq-1$ and owing to
\begin{equation}
\frac{s}{m+1}\left(V_{\mu} - A_{\mu}\right) \rightarrow \frac{s}{m+1}\left(V_{\mu} - A_{\mu}\right) + \tilde{\Omega}_{; \mu},
\end{equation}
we can finally define the "properly normalized", 4D Weyl gauge vector:
\begin{equation}
k_{\mu} \equiv \frac{s}{m+1}\left(V_{\mu} - A_{\mu}\right) \quad \text{with} \quad k_{\mu} \rightarrow k_{\mu} + \tilde{\Omega}_{; \mu} ~.
\end{equation}
This leads to the Maxwell/Weyl 4D gauge vector diagonalization which is summarized in the following table.
\begin{center}
\begin{tabular}{ |c|c|c| } 
 \hline
 4D & Maxwell & Weyl \\
 \hline 
 Gauge transformation & $A_{\mu}\rightarrow A_{\mu} + \Lambda_{;\mu}$ & $k_{\mu}\rightarrow k_{\mu}$ \\ 
 \hline
 Scale transformation & $A_{\mu}\rightarrow A_{\mu}$ & $k_{\mu}\rightarrow k_{\mu} + \tilde{\Omega}_{;\mu}$ \\ 
 \hline
\end{tabular}
\end{center}

\section{Kaluza - Klein reduction}
We start with the 5D Weyl-Dirac action in its un-starred form (\ref{5dun}).
In this section we will omit the scalar potential term. The inclusion of a scalar potential term is a 
simple process which will be done when we discuss no-scale quantum cosmology at the mini-superspace level.\\
Recall the chosen ansatz for the metric,
\begin{equation}
G_{MN} = S^{m+1}\left(
	\begin{array}{c|c}
  S^{-1}g_{\mu \nu} + A_{\mu}A_{\nu} & A_{\mu} \\
  \hline
  A_{\nu} & 1
 \end{array}\right) ~,
\end{equation}
\begin{equation} 
G^{MN} = S^{-m}\left(
	\begin{array}{c|c}
  g^{\mu \nu} & -A^{\mu} \\
  \hline
  -A^{\nu} & A^{\mu}A_{\mu} + \frac{1}{S}
 \end{array}\right) \hspace{30pt}
\end{equation}
and the Weyl gauge vector
\begin{equation}
K_{M} = s \left(
	\begin{array}{c}
	 V_{\mu} \\ 
     \hline
        1
 \end{array}\right) ~,
\end{equation}
\begin{equation} 
 K^{M} = S^{-m} \left(
	\begin{array}{c}
	 (m+1)k^{\mu} \\ 
     \hline
        \frac{s}{S}-(m+1)k^{\mu}A_{\mu}
 \end{array}\right) ~.
\end{equation}
We divide our action into four parts,
\begin{equation}
{\cal S} = \int \left(\mathcal{L}_1 + \mathcal{L}_2 +\mathcal{L}_3 + \mathcal{L}_{kin}\right)d^5x,
\end{equation}
where we have defined
\begin{equation}
\begin{array}{c}
	\mathcal{L}_1 \equiv \phi^2 \hat{R}\sqrt{-G}~,~ \mathcal{L}_2 \equiv - 4 \omega_5 G^{MN} \phi_{;M}\phi_{;N} 
	\sqrt{-G}~, \\[7pt]
	\mathcal{L}_3 \equiv (4+3\omega_5)\left(2K^M_{;M} - 3K^MK_M\right)\phi^2 \sqrt{-G}~,\hspace{18pt}\\[7pt]
	\mathcal{L}_{kin} \equiv - \frac{1}{4} \phi^{\frac{2}{3}}K^{MN}K_{MN}\sqrt{-G} ~.\hspace{75pt}
\end{array}
\end{equation}
Before we continue our analysis we wish to make \textit{a remark about the derivatives.} In the 5D action most of the derivatives are not actually covariant derivatives
but just partial derivatives, i.e $\phi_{;M} = \phi_{,M}$. This is due to the fact the they are either acting on 
scalar fields or the anti-symmetry field strength tensor. We will be examining the Kaluza-Klein vacuum, thus all 5D partial derivatives are in fact identical to their 4D counter-parts.
The only exception is the covariant derivative of the 5D Weyl gauge vector, $K^M_{;M}$. During the reduction
process this term is dealt with carefully, ensuring that after the reduction all derivatives written are their
4D variant. For the remainder of this chapter all derivatives will be their 4D variants.\\
Beginning with $\mathcal{L}_1$, for which, via a straightforward albeit lengthy calculation, the KK reduction yields,up to a total derivative,
\begin{equation}
\begin{array}{c}
\displaystyle{\mathcal{L}_1 = \left(\phi^2 S^{\frac{1}{2}+\frac{3m}{2}}R-\phi^2 S^{\frac{3}{2}+\frac{3m}{2}}\frac{1}{4}F^{\mu \nu}F_
{\mu \nu} \right.} \hspace{25mm}\\[5pt]
\displaystyle{+\frac{3}{2}m(1+2m)\phi^2 S^{-\frac{3}{2}+\frac{3m}{2}}g^{\mu \nu}S_{;\mu}S_{;\nu} \hspace{10mm}}\\[5pt]
\displaystyle{ \left.+(1+4m)S^{-\frac{1}{2}+\frac{3m}{2}}g^{\mu \nu} S_{;\mu} \phi^2_{;\nu}\right)\sqrt{-g} ~.}\hspace{8mm}
\end{array}
\end{equation}

Here $R$ is the 4D Ricci scalar and $F_{\mu \nu}$ is the $A_{\mu}$ field strength tensor.
Next are the terms $\mathcal{L}_2$ and $\mathcal{L}_3$ for which the KK reduction yields
\begin{equation}
\begin{array}{c}
\displaystyle{\mathcal{L}_2 = -4\omega_5 S^{\frac{1+3m}{2}}g^{\mu \nu}\phi_{;\mu}\phi_{;\nu}\sqrt{-g} ~,\hspace{110pt}}\\[7pt]
\displaystyle{\mathcal{L}_3 = \phi^2S^{\frac{1+3m}{2}}\sqrt{-g}(4+3\omega_5)(m+1) \times} \hspace{90pt}\\
\displaystyle{\Biggl[2k^{\mu}_{;\mu}- 3(m+1)k^{\mu}k_{\mu} +(1+3m)S^{-1}k^{\mu}S_{;\mu} - \frac{3S^{-1}s^2}{m+1}\Biggr] ~.}
\end{array}
\end{equation}

Furthermore, $\omega_5$ has to be replaced with its 4D counter part. In order to do this we recall (\ref{n-dim}) in 4D,
\begin{equation}
\begin{array}{c}
\displaystyle{	\phi^2 R^* - 4 \omega_4 g^{\mu \nu} \phi_{*\mu}\phi_{*\nu}}= \hspace{180pt} \\[3pt]
	\hspace{40pt} \phi^2 R - 4 \omega_4 g^{\mu \nu} \phi_{;\mu}\phi_{;\nu} -2\omega\left(K^{\mu} \phi^2\right)_{;\mu} \\[3pt]
	\hspace{30pt}+ (3+2\omega_4)\left(2K^{\mu}_{;\mu} - 2 K^{\mu}K_{\mu}\right)\phi^2 ~.
\end{array}
\end{equation}
Allowing us to identify the dictionary between $\omega_5$ and $\omega_4$ 
\begin{equation}
(4+3\omega_5)(m+1) = (3+2\omega_4) \quad \rightarrow \quad \omega_5 = \frac{3+2\omega_4}{3(m+1)}-\frac{4}{3} ~.
\end{equation}
To verify self-consistency, we check the $\omega_4 = -3/2$ critical case. This yields $\omega_5=-4/3$
just as it should be. Using this found dictionary we finally get
\begin{widetext}
\begin{flalign}
& \mathcal{L}_3= S^{\frac{1+3m}{2}}\phi^2 \sqrt{-g}(3+2\omega_4)\left(2k^{\mu}_{;\mu} - 3(m+1)k^{\mu}k_
{\mu}\right) \hspace{0em} \nonumber \\
& \hspace{1em}+ \phi^2 S^{\frac{1+3m}{2}}\sqrt{-g}(3+2\omega_4)\left((1+3m)S^{-1}k^{\mu}S_{;\mu} - \frac{3S^{-1}s^2}{m+1} -
\frac{4(1+4m-2\omega_4)}{3(3+2\omega_4)(m+1)}g^{\mu \nu}\phi_{;\mu}\phi_{;\nu}\right) ~.
\end{flalign}
\end{widetext}
Last but not least is the kinetic term $\mathcal{L}_{kin}$, 
\begin{equation}
\begin{array}{c}
\displaystyle{\frac{1}{4} \phi^{\frac{2}{3}}K^{MN}K_{MN}\sqrt{-G} = }\hspace{15em}\\
\displaystyle{\frac{1}{4} \phi^{\frac{2}{3}} S^{\frac{1+m}{2}} \Biggl( g^{\mu \lambda}g^{\nu \sigma}X_{\lambda \sigma} 
X_{\mu \nu} -2\left(A^{\nu}X^{\mu}_5-A^{\mu}X^{\nu}_5\right)X_{\mu \nu}}\\
\displaystyle{\hspace{1em} +2 A^{\mu}A^{\sigma}X_{\mu 5}X_{5\sigma} +2\left(A^{\mu}A_{\mu} + \frac{1}{S}\right)g^{\nu \sigma}X_{5\sigma}X_{5\nu}\Biggr)\sqrt{-g} } ~.
\end{array}
\end{equation}

This can be recast into
\begin{align}
	&\mathcal{L}_{kin}\sqrt{-G}= \nonumber\\
	&\frac{1}{4} \left(\phi S^{\frac{3(1+m)}{4}}\right)^{\frac{2}{3}}\left((m+1)k^{\mu \nu} + 
sF^{\mu \nu}\right)\left((m+1)k_{\mu \nu}+  sF_{\mu \nu}\right) \nonumber \\
& \hspace{45pt} + \frac{1}{4} \left(\phi S^{\frac{3(1+m)}{4}}\right)
^{\frac{2}{3}}\frac{2}{S}g^{\mu \sigma}s_{;\sigma}s_{;\nu} ~.
\end{align}
We have denoted the field strength tensor for the Weyl gauge vector as $k_{\mu \nu}$. In this form the Weyl-Maxwell
mixing is hard to miss. The mixing term $\frac{1}{4} \left(\phi S^{\frac{3(1+m)}{4}}\right)^{\frac{2}{3}}s(m+1)k^{\mu \nu}
F_{\mu \nu}$ is coupled only to in-scalars, who do not undergo transformations, making this term un-gaugable.\\
Finally we integrate over $x_5$ and write the final form of the reduced 4D action, up to a total derivative, in it's full 
glory
\begin{widetext}
\begin{flalign}
&{\cal S}=l_{KK} \int d^4x \sqrt{-g} \Biggl( \phi^2 S^{\frac{1}{2}+\frac{3m}{2}} R -\phi^2 S^{\frac{3}{2}+\frac{3m}{2}}\frac{1}{4}F^
{\mu \nu}F_{\mu \nu} +\frac{3}{2}m(1+2m)\phi^2 S^{-\frac{3}{2}+\frac{3m}{2}}g^{\mu \nu}S_{;\mu}S_{;\nu}  \nonumber \\
& \hspace{1em} + (1+4m)S^{-\frac{1}{2}+\frac{3m}{2}}g^{\mu \nu} S_{;\mu} \phi^2_{;\nu} +  S^{\frac{1+3m}{2}}\phi^2 (3+2\omega_4)\left
(2k^{\mu}_{;\mu} - 3(m+1)k^{\mu}k_{\mu}\right) \nonumber \\[3pt]
& \hspace{0.9em}\left.+ \phi^2 S^{\frac{1+3m}{2}}(3+2\omega_4)\left((1+3m)S^{-1}k^{\mu}S_{;\mu} - \frac{3S^{-1}s^2}{m+1} -\frac{4(1+4m
-2\omega_4)}{3(3+2\omega_4)(m+1)}\frac{g^{\mu \nu}}{\phi^2}\phi_{;\mu}\phi_{;\nu}\right) \right. \nonumber \\
& \hspace{1em} + \frac{1}{4} \left(\phi S^{\frac{3(1+m)}{4}}\right)^{\frac{2}{3}}\left((m+1)k^{\mu \nu} + sF^{\mu \nu}\right)
\left((m+1)k_{\mu \nu} + sF_{\mu \nu}\right) + \frac{1}{4} \left(\phi S^{\frac{3(1+m)}{4}}\right)^{\frac{2}{3}}\frac{2}{S}
g^{\mu \sigma}s_{;\sigma}s_{;\nu} \Biggr) ~.
\end{flalign}
\end{widetext}
We make a few remarks concerning our reduced action,
\begin{itemize}
\item If one wishes that the action have no ghost terms $m$ can not be completely arbitrary. The strongest $m$ constraint is due to the $S$ field kinetic term coefficient where it must be that $m(1+2m)<0 ~ \Rightarrow ~ -\frac{1}{2} < m < 0$.
\item Assuming $x_5$ - independence at the Kaluza-Klein vacuum level, we can trivially integrate out the fifth dimension.
\item The classical 4D equations of motion must be local scale invariant (not sensitive to $l_{KK}$).
\item The higher Kaluza-Klein fourier modes $~e^{\pm i \frac{p x_5}{l_{kk}}}$, where $p=1,2,...$, are already $l_{KK}$ - dependent.
\end{itemize}

\section{The $m=-\frac{1}{3}$ case}
An interesting case to examine is the $m=-\frac{1}{3}$ one. First we note the elegant $\omega_4$ to $\omega_5$ dictionary,
\begin{equation}
\omega_4 = \omega_5 - \frac{1}{6} ~.
\end{equation}
Secondly we denote the Kaluza-Klein radius in a more convenient manner - $l\equiv l_{KK}$.\\
For this case the action, 
\begin{equation}
\begin{array}{c}
\displaystyle{{\cal S}=l \int \sqrt{-g} d^4x \Biggl( \phi^2 R -\phi^2 S\frac{1}{4}F^{\mu \nu}F_{\mu \nu} }\hspace{8em}\\
\displaystyle{-\frac{1}{6}\phi^2 S^{-2}g^{\mu \nu}S_{;\mu}S_{;\nu} + \frac{1}{4} \left(\phi S^{\frac{1}{2}}\right)^{\frac{2}{3}}\frac{2}{S}g^{\mu \sigma}s_{;\sigma}s_{;\nu}\hspace{1.5em}}\\[10pt]
\displaystyle{\hspace{1.3em}- \frac{2}{3}(1+ 6\omega_4)g^{\mu \nu}\phi_{;\mu}\phi_{;\nu}+  2\phi^2 (3+2\omega_4)\left(k^{\mu}_{;\mu} - k^{\mu}k_{\mu}\right)}\\[5pt]
\displaystyle{ - \phi^2 (3+2\omega_4)\frac{9}{2}S^{-1}s^2 -\frac{1}{3}S^{-1}g^{\mu \nu} S_{;\mu} \phi^2_{;\nu} \hspace{3.8em}}\\[3pt]
\displaystyle{\hspace{2.5em} + \frac{1}{4} \left(\phi S^{\frac{1}{2}}\right)^{\frac{2}{3}}\left(\frac{2}{3}k^{\mu \nu} + sF^{\mu \nu}\right)\left(\frac{2}{3}k_{\mu \nu} + sF_{\mu \nu}\right)\Biggr) }~,\hspace{2.2em}
\end{array}
\end{equation}
features no coupling between $S$ and the curvature.\\
It is possible, yet not very informative, to star this action. The starring will leave us with a simpler action but with the above mentioned trade-off, we do not have physical intuition about it.
Additionally we can define a new variable $\sigma = \phi^2 S$. Finally we write the action in the terms of two in-scalars and one co-scalar
\begin{equation}
\begin{array}{c}
\displaystyle{{\cal S} = l \int \sqrt{-g} d^4x  \Biggl( \phi^2 R^* - 4 \omega_4g^{\mu \nu} \phi_{*\mu}\phi_{*\nu}  - \frac{1}{4} \sigma F^{\mu \nu}F_{\mu \nu} }\\
\displaystyle{- \frac{1}{6}\phi^2 g^{\mu \nu} \frac{\sigma_{*\mu}}{\sigma}\frac{\sigma_{*\nu}}{\sigma}-\frac{9}{2}(3+2\omega_4) \frac{s^2}{\sigma^2} \phi^4 + \frac{1}{2} \frac{\phi^2}{\sigma^{\frac{2}{3}}} g^{\mu \nu}s_{*\mu}s_{*\nu}} \\
\hspace{6em}\displaystyle{ + \frac{1}{4}\sigma^{\frac{1}{3}}\left(\frac{2}{3}k^{\mu \nu} + sF^{\mu \nu}\right)\left(\frac{2}{3}k_{\mu \nu} + sF_{\mu \nu}\right)  \Biggr)} 
\end{array}
\end{equation}
with the transformation rules
\begin{eqnarray}
\sigma \rightarrow \sigma \quad ; \quad s \rightarrow s\quad ; \quad \phi \rightarrow e^{-\Omega}\phi
\end{eqnarray}
Unit wise - our action is not composed of "normal" scalar (with length unit $L^{-1}$).
In-addition, following our ansatz, all vector fields are also not "normal" (for further details see table \ref{oldvar}).
\begin{table}
  \begin{tabular}{| c | c | c |}
    \hline
    5D  &  Scale transformation  &  Units\\ \hline
    $x^{M}$ & $x^M$  &  $L^1$ \\ \hline 
    $\sqrt{-G}$ & $e^{-5\Omega}\sqrt{-G}$ & $L^0$ \\ \hline
    $G_{MN}$ & $e^{-2\Omega}G_{MN}$ & $L^0$ \\ \hline
    $G^{MN}$ & $e^{2\Omega}G^{MN}$ & $L^0$ \\ \hline
    $R^*_{MN}$ & $R^*_{MN}$ & $L^{-2}$ \\ \hline
    $R^*$ & $e^{2\Omega} R^*$ & $L^{-2}$ \\ \hline
    $\phi$ & $e^{\frac{3}{2}\Omega}\phi$ & $L^{-\frac{3}{2}}$ \\ \hline
    $\phi_{*M}$ & $e^{\frac{3}{2}\Omega}\phi_{*M}$ & $L^{-\frac{5}{2}}$ \\ \hline
    $K_M$ & $K_M + \frac{3}{2}\Omega_{,M}$ & $L^{-1}$ \\ \hline
    $K_{MN}$ & $K_{MN}$ & $L^{-2}$ \\ \hline
  \end{tabular}
  
  \begin{tabular}{| c | c | c |}
    \hline
    4D  &  Scale transformation  &  Units\\ \hline
    $x^{\mu}$ & $x^{\mu}$  &  $L^1$ \\ \hline 
    $\sqrt{-g}$ & $e^{-4\tilde{\Omega}}\sqrt{-g}$ & $L^0$ \\ \hline
    $g_{\mu \nu}$ & $e^{-2\tilde{\Omega}}g_{\mu \nu}$ & $L^0$ \\ \hline
    $g^{\mu \nu}$ & $e^{2\tilde{\Omega}}g^{\mu \nu}$ & $L^0$ \\ \hline
    $A_{\mu}$ & $A_{\mu}$ & $L^{-1}$ \\ \hline
    $S$ & $e^{-2 \tilde{\Omega}}S$ & $L^0$ \\ \hline
    $R^*_{\mu \nu}$ & $R^*_{\mu \nu}$ & $L^{-2}$ \\ \hline
    $R^*$ & $e^{2\tilde{\Omega}} R^*$ & $L^{-2}$ \\ \hline
    $\phi$ & $e^{\tilde{\Omega}}\phi$ & $L^{-\frac{3}{2}}$ \\ \hline
    $\phi_{*\mu}$ & $e^{\tilde{\Omega}}\phi_{*\mu}$ & $L^{-\frac{5}{2}}$ \\ \hline
    $s$ & $s$ & $L^{-1}$ \\ \hline
    $k_{\mu}$ & $k_{\mu} + \tilde{\Omega}_{,\mu}$ & $L^{-1}$ \\ \hline
    $k_{MN}$ & $k_{MN}$ & $L^{-2}$ \\ \hline
  \end{tabular}
  \caption{Length and scale transformations for 5D and 4D variables}
  \label{oldvar}
\end{table}
This stems from the reduction where the 5D length length scales are set to make sure
that the action is length-less. For proper length units we must add a length scale, the Kaluza-Klein radius,
into the ansatz
\begin{equation}
		G_{MN} = S^{m+1}\left(
	\begin{array}{c|c}
  S^{-1}g_{\mu \nu} +l^2 A_{\mu}A_{\nu} & lA_{\mu} \\
  \hline
  lA_{\nu} & 1
 \end{array}\right) ~,
\end{equation}
\begin{equation}
K_{M} = s \left(
	\begin{array}{c}
	 l V_{\mu} \\ 
     \hline
        1
 \end{array}\right) ~.
	\end{equation}
This leads to the reduced action,	
\begin{equation}
\begin{array}{c}
\displaystyle{{\cal S} = l \int  \left( \phi^2 R^* - 4 \omega_4g^{\mu \nu} \phi_{*\mu}\phi_{*\nu}  - \frac{l^2}{4} \sigma F^{\mu \nu}F_{\mu \nu}\right.}\hspace{4em}\\
\displaystyle{ - \frac{1}{6}\phi^2 g^{\mu \nu} \frac{\sigma_{*\mu}}{\sigma}\frac{\sigma_{*\nu}}{\sigma}-\frac{9}{2}(3+2\omega_4) \frac{s^2}{\sigma^2} \phi^4 + \frac{1}{2} \frac{\phi^2}{\sigma^{\frac{2}{3}}} g^{\mu \nu}s_{*\mu}s_{*\nu}}\vspace{4pt}\\
\displaystyle{\left. + \frac{l^2}{4} \sigma^{\frac{1}{3}}\left(\frac{2}{3}k^{\mu \nu} + sF^{\mu \nu}\right)\left(\frac{2}{3}k_{\mu \nu} + sF_{\mu \nu}\right)  \right)\sqrt{-g} d^4x} ~.\hspace{10pt}
\end{array}
\end{equation}
We now redefine our scalar fields as follows,
\begin{equation}
s \rightarrow l^A s, \quad \phi\rightarrow l^B\phi, \quad S\rightarrow l^C S,
\end{equation}
allowing us to find the proper powers which leave the action $l$ - independent.
Let us look at following terms,
\begin{flalign}
&\frac{l^3}{4} \sigma F^{\mu \nu}F_{\mu \nu}\rightarrow \frac{1}{4} l^{3+C+2B}\sigma F^{\mu \nu}F_{\mu \nu} \\
&l \phi^2 R^* \rightarrow l^{1+2B} \phi^2 R^* \\
&l\frac{9}{2}(3+2\omega_4) \frac{s^2}{\sigma^2} \phi^4 \rightarrow \frac{9}{2}(3+2\omega_4) l^{1+2A-C+2B} \frac{s^2}{\sigma} \phi^4 ~.
\end{flalign}
This leads to
\begin{equation}
\begin{array}{c}
3+C+2B=0~,\\
1+2B=0 ~,\\
1+2A-C+2B=0 ~.
\end{array}
\end{equation}
Solving these we find $A=-1$, $B=-\frac{1}{2}$, and $C=-2$.
\begin{table}
  \begin{tabular}{| c | c | c |}
    \hline
    4D  &  Scale transformation  &  Units\\ \hline
    $s$ & $s$ &  $L^0$ \\ \hline 
    $\phi$ & $e^{\tilde{\Omega}}\phi$ & $L^{-1}$ \\ \hline
    $S$ & $e^{-2\tilde{\Omega}}S$ & $L^2$ \\ \hline
    $\sigma$ & $\sigma$ & $L^0$ \\ \hline
  \end{tabular}
  \caption{Length and scale transformations for the newly defined 4D variables}
  \label{newvar}
\end{table} 
We can finally write the reduced, diagonalized, action
\begin{flalign}
\begin{array}{c}
\displaystyle{\mathcal{S} = \int \sqrt{-g} d^4x \Biggl( \phi^2 R^* - 4 \omega_4g^{\mu \nu} \phi_{*\mu}\phi_{*\nu}- \frac{1}{4} \sigma F^{\mu \nu}F_{\mu \nu}}  \hspace{3em} \\
 \displaystyle{- \frac{1}{6}\phi^2 g^{\mu \nu} \frac{\sigma_{*\mu}}{\sigma}\frac{\sigma_{*\nu}}{\sigma}-\frac{9}{2}(3+2\omega_4) \frac{s^2}{\sigma^2} \phi^4 + \frac{1}{2} \frac{\phi^2}{\sigma^{\frac{2}{3}}} g^{\mu \nu}s_{*\mu}s_{*\nu}}  \\[10pt]
 \displaystyle{ + \frac{1}{4} \left(\frac{2}{3}k^{\mu \nu} + sF^{\mu \nu}\right)\left(\frac{2}{3}k_{\mu \nu} + sF_{\mu \nu}\right)\sigma^{\frac{1}{3}} \Biggr)}~.\hspace{50pt}
\end{array}
\end{flalign}
With the KK radius properly absorbed into the scalar fields the action is composed of
two in-scalars ($\sigma$ and $s$) and one co-scalar ($\phi$). Length-wise each in-scalar
is unit-less as it brings no scale into the action, as such they are not "normal" scalar 
fields with the proper $L^{-1}$ length unit (see table \ref{newvar}). On the other hand the co-scalar has the 
proper scalar field length units, this invites the idea, that a mass term may emerge from the vacuum expectation value of this field. 
\section{No-scale Kaluza-Klein quantum cosmology}
\begin{figure}[H]
	\center
	\includegraphics[width=.45\linewidth, scale=1]{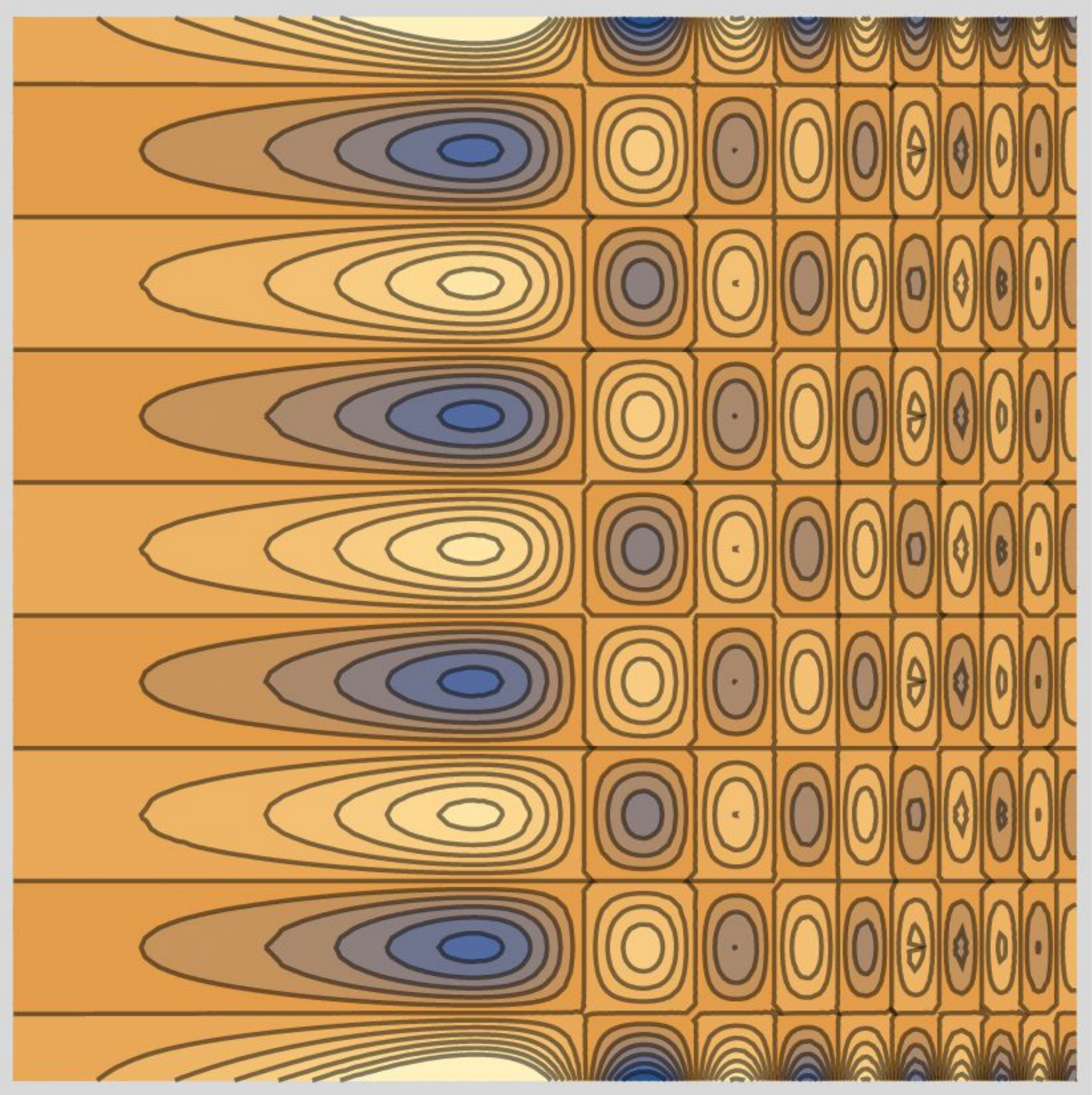} ~
	\includegraphics[width=.45\linewidth, height= 38.9mm]{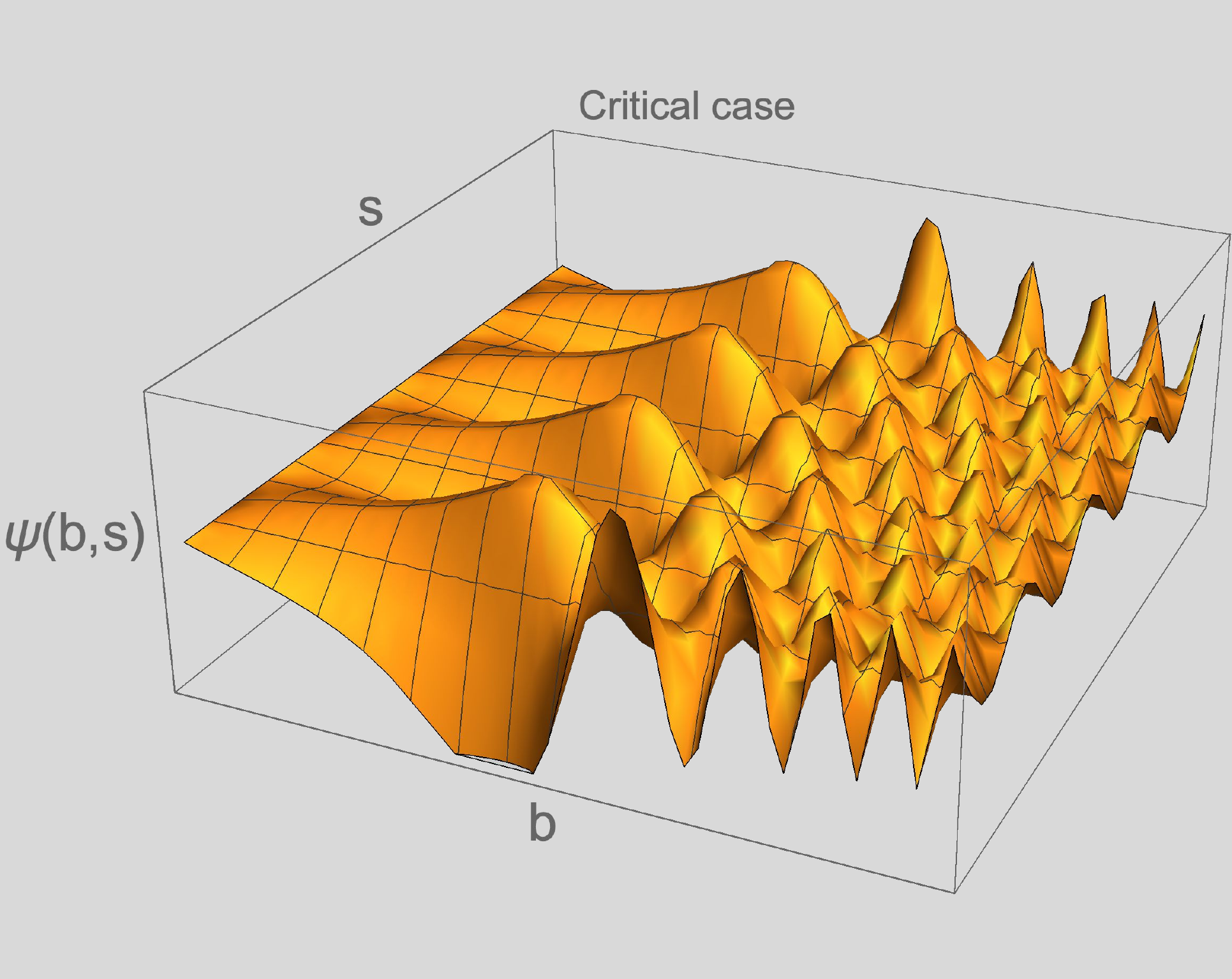}	
	\caption{The 3D (right) and contour (left) plots of the
	critical, namely $2\omega_4 + 3 =0$, no-scale cosmological wave
	function $\psi(b,s)$, which is subjected to the automatic deWitt
	initial condition $\psi(0,s)=0$. This case is plotted for $\eta<0$, which strongly reminds us (up to $s$-periodicity) of the
	Hartle-Hawking solution.}
	\label{critical}
\end{figure}

\begin{figure}[H]
	\center
	\includegraphics[width=.45\linewidth, scale=1]{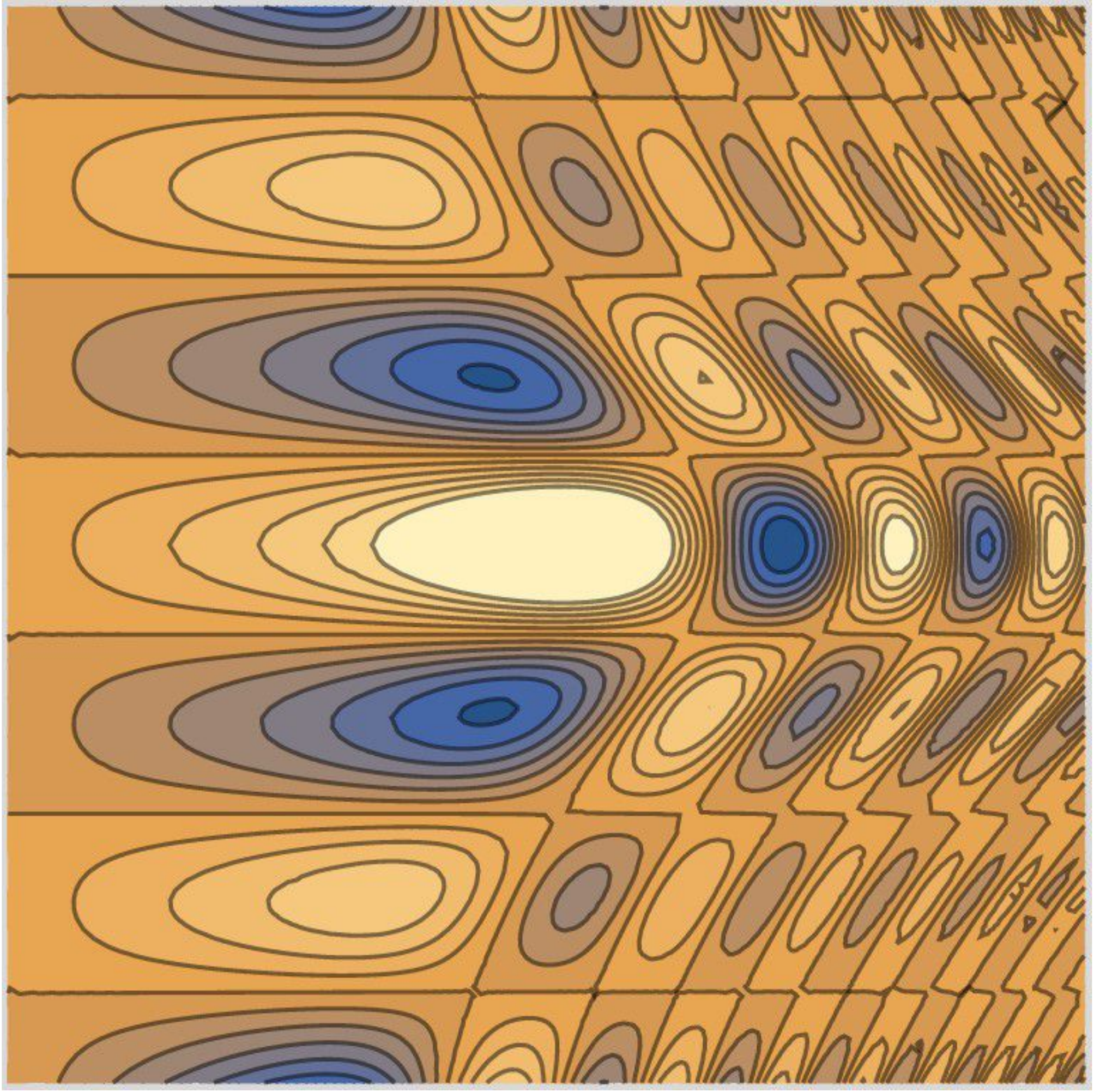}~~
	\includegraphics[width=.45\linewidth, height= 38.9mm]{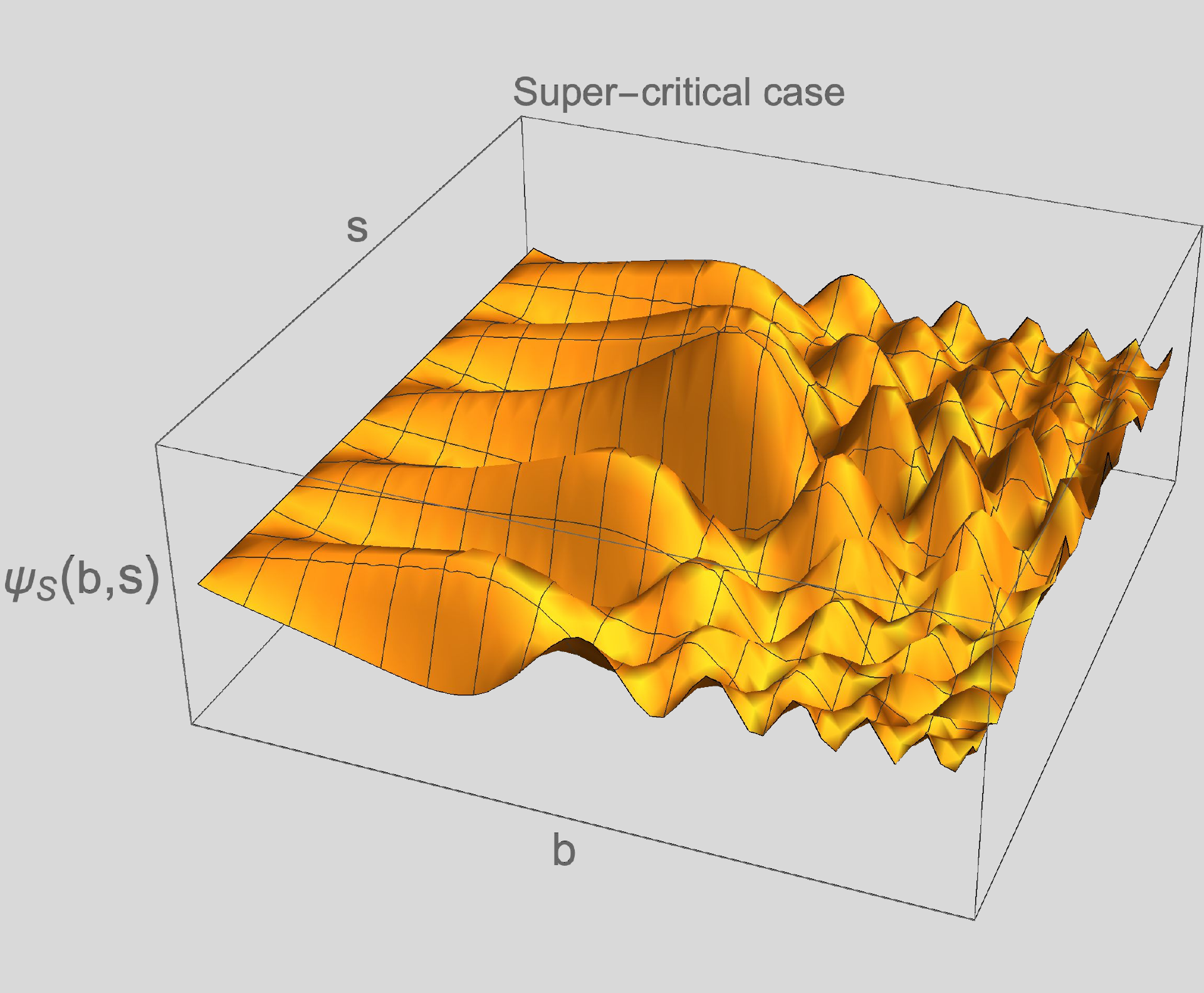}	
	\caption{The 3D (right) and contour (left) plots of the super 
	critical, namely $2\omega_4 + 3 >0$, no-scale cosmological wave
	function $\psi(b,s)$, which is subjected to the automatic deWitt
	initial condition $\psi(0,s)=0$. In addition to surviving the $\Lambda\rightarrow 0$ limit, this case favors the small values of $s$.}
	\label{super_critical}
\end{figure}

Our starting point is the $m=-1/3$ reduced action. If we wish to follow Hartle and Hawking our action must include a scalar potential term. It must also stem from the 5D scalar potential (Recall $V(\phi)= -2\Lambda\phi^{10/3}$), which is trivially reduced to $V(\phi) = - \frac{2 \Lambda}{\left(\phi^2 S\right)^{1/3}}\phi^4$. Plugging this into our action
\begin{equation}
\begin{array}{c}
\displaystyle{\mathcal{S} = \int  \Biggl( \phi^2 R^*_4 - 4 \omega_4g^{\mu \nu} \phi_{*\mu}\phi_{*\nu}  - \frac{1}{4} \sigma F^{\mu \nu}F_{\mu \nu}} \vspace{4pt}\\
 \displaystyle{- \frac{1}{6}\phi^2 g^{\mu \nu} \frac{\sigma_{*\mu}}{\sigma}\frac{\sigma_{*\nu}}{\sigma}+ \frac{1}{2} \frac{\phi^2}{\sigma^{\frac{2}{3}}} g^{\mu \nu}s_{*\mu}s_{*\nu}} \vspace{4pt} \\ 
\hspace{2em} \displaystyle{+ \frac{1}{4} \sigma^{\frac{1}{3}}\left(\frac{2}{3}k^{\mu \nu} + sF^{\mu \nu}\right)\left(\frac{2}{3}k_{\mu \nu} + sF_{\mu \nu}\right)} \vspace{4pt}\\
\hspace{2.7em}\displaystyle{ -\left(\frac{9}{2}(3+2\omega_4) \frac{s^2}{\sigma^2}+\frac{2 \Lambda}{\sigma^{1/3}}\right) \phi^4 \Biggr)\sqrt{-g} d^4x ~.}
\end{array}
\end{equation}
At the mini-superspace level, cosmology can only tolerate the pure gauge configurations
\begin{equation}
	A_{\mu}=(A(t),0,0,0) ~,~~  k_{\mu}=(v(t),0,0,0) ~.
\end{equation}
As a result, $F_{\mu\nu}=k_{\mu\nu}=0$ leaving the extraordinary Weyl-Maxwell mixing out of
the game. At the moment we will abandon our $\sigma$ notation, leaving us with three
scalar fields $s, S, \phi$. Following the standard procedure of integrating out over the maximally symmetric space
\begin{equation}
\int\mathcal{L}\sqrt{-g}dt d^3x ~\rightarrow ~ \int\mathcal{L}_{mini}dt ~,
\end{equation}
and up to a total derivative, the mini superspace Lagrangian is
\begin{equation}
\begin{array}{c}
	\displaystyle{\mathcal{L}_{mini} = \frac{6 a \phi^2 \dot{a}^2 }{n} + \frac{12 a^2 \phi \dot{\phi} \dot{a}}{n}-6\kappa n a \phi^2 -\frac{a^3 \phi2 \dot{S}^2}{6 n S^2}} \hspace{60pt}\\[5pt]
	\displaystyle{- \frac{2 a^3 \phi \dot{S} \dot{\phi}}{3 n S} - \frac{2 a^3(1+6\omega_4)\dot{\phi}^2}{3n}-2(3+2\omega_4)\frac{a^3 v^2 \phi^2}{n}}\hspace{5pt}\\[5pt]
	\displaystyle{-(3+2\omega_4)\frac{4 a^3 \phi \dot{\phi} v}{n}+ \frac{9}{2}(3+2\omega_4)\frac{a^3 n \phi^2 s^2}{S}} \hspace{38pt}\\[5pt]
\displaystyle{+ \frac{\Lambda a^3 n \phi^{10/3}}{S^{1/3}} - \frac{a^3 \dot{s}^2 \phi^{2/3}}{2 n S^{2/3}}} ~,\hspace{105pt}
\end{array}
\end{equation}
while reviving the lapse function $n(t)$ to keep track of the
underlying diffeomorphism.
We translate the mini-superspace Lagrangian to the
Hamiltonian formalism. Using the Legendre transform $H = p_i\dot{q}_i - L $ and the cannonical momentum
\begin{subequations}
\begin{flalign}
&p_a = \frac{12 a \phi^2 \dot{a}}{n} + \frac{12 a^2 \phi \dot{\phi}}{n}, \\
&p_{\phi} = \frac{12 a^2 \phi \dot{a}}{n} - \frac{2 a^3 \phi \dot{S}}{3 n S} - \frac{4 a^3 (1+6\omega_4) \dot{\phi}}{3 n} \nonumber\\
& \hspace{12pt}- (3+2\omega_4)\frac{4 a^3 \phi ^2 v}{n}, \\
&p_S = - \frac{2 a^3 \phi^2 \dot{S}}{6 n S^2}, \\
&p_s = -\frac{a^3 \dot{s} \phi^{2/3}}{n S^{2/3}} ~.
\end{flalign}
\end{subequations}
The resulting Hamiltonian,
\begin{equation}
\begin{array}{c}
	{\cal H}=v(a p_a-\phi p_\phi+2S p_S) \hspace{120pt}\\[3pt]
	\displaystyle{-\frac{n}{a}\left[ 
	\frac{p_a^2}{24\phi^2}
	-\frac{3S^2 p_S^2}{2a^2\phi^2}
	-\frac{S^{\frac{2}{3}}p_s^2}{2a^2 \phi^{\frac{2}{3}}}
	-\frac{(a p_a-\phi p_\phi+2S p_S)^2}{8(3+2\omega)a^2\phi^2}
	\right.}\\[5pt]
	\hspace{40pt}\displaystyle{\left. +6\kappa a^2\phi^2
	-\left(\frac{9(3+2\omega_4)s^2}{2S\phi^2}
	+\frac{2\Lambda}{S^{\frac{1}{3}}\phi^{\frac{2}{3}}}
	\right)a^4\phi^4\right]} ~,
	\label{HKK}
\end{array}
\end{equation}
is linear in $v ,n$. This gives rise to two first class constraints.
We are mainly interested in the a quantum no-scale
cosmology. Thus we will proceed directly to the pair of Schrodinger equations, 
skipping the classical equations of motion. We immediately recognize the coefficient
of $v$ in Eq.(\ref{HKK}) as the $\hbar$-independent scale symmetry constraint, leading to
\begin{equation}
	\left(a\frac{\partial}{\partial a}
	-\phi\frac{\partial}{\partial \phi}
	+2 S\frac{\partial}{\partial S}
	\right)\psi(a,\phi,S,s)=0 ~.
\end{equation}
This constraint forces the wave function to depend solely on
Dirac's in-scalars,
\begin{equation}
	a^\alpha \phi^\beta S^\gamma
	~~\quad \text{with}~ \alpha-\beta+2\gamma=0 ~.
\end{equation}
Without losing generality, the simplest choice
would be
\begin{equation}
	\psi(a,\phi,S,s)=\psi(a\phi,log\left(S\phi^2\right),s) ~,
\end{equation}
for which we can now use the short hand in-scalar notations
\begin{equation}
	b=a\phi~,~~ z=\log S\phi^2 =\log \sigma~.
\end{equation}
Note how $\sigma$ re-enters the theory, as was anticipated, being a fundamental building block of the theory.
The associated Hamiltonian constraint, identified as the
coefficient of $n$ in Eq.(\ref{HKK}), eventually becomes
the zero energy Schrodinger equation
\begin{equation}
	-\frac{\hbar^2}{24}
	\frac{\partial^2 \psi}{\partial b^2}
	+\frac{3\hbar^2}{2b^2}
	\frac{\partial^2 \psi}{\partial z^2}
	+\frac{e^{\frac{2}{3}z}\hbar^2}{2b^2}
	\frac{\partial^2 \psi}{\partial s^2}
	+V(b,z,s)\psi=0 ~,
\end{equation}
with the accompanying potential being
\begin{equation}
	V(b,z,s)=6\kappa b^2
	-\left(\frac{9}{2}(3+2\omega_4)e^{-z}s^2
	+2e^{-\frac{1}{3}z}\Lambda \right)b^4 ~.
\end{equation}
In some respects, the $b^4$ coefficient can be thought to be
taking the part of $\Lambda_{eff}(s, \Lambda)$. It is quite
remarkable how even in the case of $\Lambda=0$ the $b^4$ coefficient,
provided $3+2\omega_4>0$, can still be positive.
The door is now widely open for a variety of spacial cases. On
simplicity grounds, we choose to analyze the case of a constant Kaluza-
Klein in-radius. In the standard Kaluza-Klein theory, with the line 
element $ds_5^2=S^{-\frac{1}{3}}ds_4^2+S^{\frac{2}{3}}\ell^2(d\theta+A_\mu dx^\mu)^2$,
the invariant 5D radius is given by $S^{\frac{1}{3}}\ell$. However
in the original works of Kaluza and Klein the radius was chosen to 
be constant, $S=1$, in order for the theory to resemble the Einstein-Maxwell action.
Unfortunately, such a choice does not make any sense once local scale symmetry is
applied. Instead, the closest choice one can make is by freezing a tenable 
in-scalar (which has nothing to do with gauge fixing), with the most obvious
choice being
\begin{equation}
	\sigma=1 ~.
\end{equation}
This will allow us to focus on the special role played by the Weyl 4D
in-scalar $s$. The corresponding wave function, $\psi(b,s)$, obeys the 
Hartle-Hawking equation
\begin{eqnarray}
	&\displaystyle{-\frac{\hbar^2}{24}
	\frac{\partial^2 \psi}{\partial b^2}
	+\frac{\hbar^2}{2b^2}
	\frac{\partial^2 \psi}{\partial s^2}
	+V(b,s)\psi=0 ~,}& \\
	& \displaystyle{V(b,s)=6\kappa b^2
	-\left(\frac{9}{2}(3+2\omega_4)s^2
	+2\Lambda \right)b^4} ~.&
\end{eqnarray}
We begin our analysis by looking at the critical case $\omega_4=-\frac{3}{2}$.
This is the simplest choice we can make as it allows for the separation of
variables $\psi(b,s)=f(b)g(s)$. $f(b)$ serves as a modified Hartle-Hawking
wave function subject to the effective potential
\begin{equation}
	V_{eff}(b)=\frac{\eta\hbar^2}{2b^2}
	+6\kappa b^2-2\Lambda b^4~.
\end{equation}
The constant $\eta$ governs the equation
\begin{equation}
	g^{\prime\prime}(s)=\eta g(s) ~.
\end{equation}
The sign of $\eta$ dictates the behavior of our theory. Especially 
interesting is the behavior near the Big Bang origin $b\rightarrow 0$.
There are three possible cases:
\begin{enumerate}
\item $\eta=0$ - yields a fully recovered Hartle-Hawking model,
accompanied by $g(s)=const$. In addition the no-boundary proposal is
recovered for $f(b)\sim b$.

\item $\eta>0$ - yields an unbounded $g(s)=e^{\pm\sqrt{\eta}s}$, 
which might imply a non-physical case.

\item $\eta<0$ - yields a well behaved $g(s)=e^{\pm i \sqrt{|\eta|}s}$.
Furthermore, if $\eta\hbar^2 \Lambda^2+16\kappa^3>0$ the cosmic evolution
undergoes an embryonic era, as is evident in this case by the shape of the
effective potential. The no-boundary proposal is not recovered, however 
both solutions $f_{1,2}(b)\sim b^{\delta_{1,2}}$ (with $0<Re(\delta_{1,2})<1$)
vanish asymptotically at the origin. This is, in-fact, the deWitt initial condition
emerging from the theory automatically.
\end{enumerate}
For further details, see Fig.(\ref{critical}).
In comparison, the non-critical case is characterized by an effective cosmological constant
\begin{equation}
	\Lambda_{eff}(s)=\Lambda+\frac{9}{4}(3+2\omega_4)s^2 ~.
\end{equation}
The additional term is positive for a super-critical 
Brans-Dicke parameter $\omega_4 >-\frac{3}{2}$ (including, in particular,
the ghost-free case $\omega_4 \geq 0$). Sadly the separation of variables
method does not work any more. However, the structure of the Schrodinger 
equation makes the value of $\omega_4$ irrelevant to the behavior of the 
wave function near the Big Bang. The larger $\omega_4$, the more concentrated
is the wave function around $s^2\ll 1$.
For further details, see Fig.(\ref{super_critical}).

\acknowledgments
{TY was supported by the Israeli Science Foundation grant $\# 1635/16$.}

\end{document}